\documentclass{article}
\usepackage{amssymb}

\begin{document}

\title{On a class of mappings between Riemannian manifolds}
\author{Thomas H. Otway\thanks{%
email: otway@yu.edu} \\
%EndAName
\\
\textit{Department of Mathematics, Yeshiva University,}\\
\textit{\ New York, New York 10033}}
\date{}
\maketitle

\begin{abstract}
Effects of geometric constraints on a steady flow potential are
described by an elliptic-hyperbolic generalization of the harmonic
map equations. Sufficient conditions are given for global
triviality. \textit{MSC2000}: 58E20, 58E99, 75N10.
\end{abstract}

\section{Introduction}

In \cite{O3}, local properties of maps which are critical points of
a nonlinear Hodge energy were investigated (see also \cite{O1}). The
target of the map has a physical interpretation as a geometric
constraint on the potential of a steady flow. In this note we
illustrate these maps by deriving elementary but explicit physical
examples, clarify the relation of the objects studied in \cite{O3}
to other classes of maps which have been studied recently, and
provide global conditions under which critical points reduce to the
trivial map.

\subsection{A column of tap water as a mapping}

A simple motivating example for placing geometric constraints on a
flow potential is provided by the steady flow of water from a
faucet. If $v_i$ is the velocity of the flow through a thin
horizontal section of area $A_i$ and if $v_f$ and $A_f$ are
defined analogously, then the conservation of mass implies that

\begin{equation}\label{conserve}
    v_iA_i=v_fA_f.
\end{equation}
But the particles accelerate under gravity, so if the cross
section $A_f$ is taken nearer to the drain than the cross section
$A_i,$ we conclude that $v_f>v_i.$ Equation (\ref{conserve}) then
implies that $A_f<A_i,$ which explains why the column of water is
seen to taper. In this conventional approach the flow geometry is
derived by imposing a physical law (acceleration under gravity) on
a conservation law.

An alternative approach would be to observe that the column of
water tapers, so that $A_f <A_i.$ Conservation of mass implies
eq.\ (\ref{conserve}), so we conclude that $v_f>v_i,$ that is, the
particles accelerate under gravity. In this alternative approach
the physical law is derived by imposing an observed flow geometry
on a conservation law.

It is convenient to think of the geometry as a constraint applied
to a flow potential $u.$ In this example $u$ maps a right circular
cylinder into a tapered cylinder. Recall that every smooth curve
has a dual representation as the envelope of its family of tangent
lines. In a steady flow the velocity vectors appear as tangent
lines to the potential surfaces, and so form an envelope of the
cross sections represented in eq.\ (\ref{conserve}). We see these
cross sections taper by tracing the velocity vectors of the water
droplets. Thus, as is often the case in fluid dynamics, the
mathematical abstraction of a potential surface attains a visible
representation in physical space.

While this simple example involves incompressible flow, the same
alternatives exist in the more complicated case of compressible
flow such as the flow of exhaust from a jet engine.

\subsection{Shallow hydrodynamic flow}

Now we consider a slightly more complicated case, that of steady,
inviscid, hydrodynamic flow in a shallow channel. Write the flow
velocity $v$ in components $\left( v_1,v_2,v_3\right) ,$ where
$v_1$ is the horizontal component in the $x$-direction, $v_2$ is
the horizontal component in the $y$-direction, and $v_3$ is the
component in the (vertical) $z$-direction. Impose initial
conditions under which $v_3$ is zero at time $t=0.$ Because we are
assuming shallow depth, it is reasonable to suppose that the
component of acceleration of water particles in the $z$-direction
has negligible effect on pressure. The result of applying this
hydrostatic law is that $v_3$ remains zero for all subsequent
times and the horizontal velocity components $v_1$ and $v_2$ are
independent of the $z$-coordinate. Because the flow is steady, the
velocity components are also independent of $t.$

Generalizing eq.\ (\ref{conserve}) to express the vanishing of an
appropriate surface integral and applying the Divergence Theorem,
we write the law of mass conservation in the form of a
\emph{continuity equation} (see, \emph{e.g.}, \cite{J}, Sec.\
1.1.1)

\begin{equation}\label{continuity1}
    \frac{\partial}{\partial x}\left[ h\left( x,y\right) v_1\left( x,y\right) %
\right] +\frac{\partial}{\partial y}\left[ h\left( x,y\right)
v_2\left( x,y\right) \right] = 0,
\end{equation}
where $h(x,y)$ represents the depth of the channel at the point
$(x,y).$ Bernoulli's formula expresses $h$ as a function of
$Q\equiv \left|v\right|^2,$ that is, $h\left( Q\right)
=\left(C-Q\right)/2g,$ where $C$ is a constant and $g$ is the
magnitude of gravitational acceleration. Substituting this
relation into (\ref{continuity1}) and using the chain rule, we
find that prior to the imposition of any geometric constraint the
flow will satisfy

\begin{equation}\label{continuity2}
    \left[ \frac{C-Q}{2}-v_1^{2}\right] v_{1x}-v_1v_2\left(v_{1y}+
v_{2x}\right) +\left[ \frac{C-Q}{2}-v_2^{2}\right]v_{2y}=0.
\end{equation}
(In eq.\ (\ref{continuity2}) numerical subscripts denote vector
components, whereas variable subscripts denote partial
differentiation in the direction of the variable.) When we applied
Bernoulli's formula we tacitly assumed the flow to be
irrotational. Thus its velocity vector has vanishing curl, which
allows us to equate mixed partial derivatives and assume the local
existence of a potential function $u(x,y)$ such that $\nabla u=v.$

Writing $c^{2}=gh=\left(C-Q\right)/2,$ we obtain (\emph{c.f.}
(10.12.5) of \cite{St}) a second-order quasilinear
elliptic-hyperbolic equation for the potential function:

\begin{equation}\label{flow}
    \left[ c^{2}-u_x^{2}\right] u_{xx}
    -2u_xu_yu_{xy}+\left[ c^{2}-u_y^{2}\right] u_{yy}=0.
\end{equation}
The type of eq.\ (\ref{flow}) depends on whether or not the flow
speed $\sqrt{Q}$ exceeds the propagation speed $c.$ For
\textit{subcritical} flow speeds in which the \emph{Froude number}
$F=\sqrt{Q}/c$ is exceeded by 1, the continuity equation is of
elliptic type and the flow is \emph{tranquil}. For
\textit{supercritical} flow speeds in which $F$ exceeds 1, eq.\
(\ref{flow}) is of hyperbolic type, which characterizes
\emph{shooting flow}.

We can prescribe flow geometry for this problem by means analogous
to the simpler example of water flowing from a faucet. In this
case the potential function can be considered as a map from the
flow domain to a target manifold. In order for the map to have
non-trivial geometry, we must assume that the potential function
$u$ is multi-valued. Its components can be imagined as local
coordinates on the target manifold. However, unlike the usual
representations of the flow potential in the complex plane
(\emph{e.g.,} Sec.\ 1.6 of \cite{Me}), our mappings are defined
over the real field. A geometric variational problem for this
model will be formulated in the next section. Note that
multi-valued potentials arise naturally on flow domains having
non-trivial topology.

Conjectures on the observable effects of geometric constraints on
the flow of shallow water go back at least 400 years. Geometric
arguments, to one degree or another, have been applied to explain
tidal bores on the English rivers Severn and Trent, the French
river Seine near Caudebec-en-Caux, and the Chinese river
Tsien-Tang, as well non-tidal anomalies such as those involving
the Agulhas Current. See \cite{O5} for a review. As our last
physical example, we recall a model for those effects which has
particularly simple geometry; the model is given in greater detail
in \cite{Tr}.

Consider a steady current flowing in the positive-$x$ direction.
Suppose that an incline of magnitude $\delta$ occurring between
the points $x_{0}$ and $x_{1}$ of an otherwise horizontal channel
floor produces a surface elevation of height $\varepsilon$ at
$x=x_{1}.$ Suppose that $x_{0}<x_{1},$ that the velocity of the
flow to the left of $x_{0}$ is $v_1$ and that the velocity of the
flow to the right of $x_{1}$ is $\widetilde{v_1}.$ An inviscid
flow is a conservative system. Equating the kinetic and potential
energies of the flow for an arbitrary surface particle of mass
$m,$ we have

\begin{equation}\label{conserve1}
    \frac{mv_1^{2}}{2}-\frac{m\widetilde{v_1}^{2}}{2}=mg\varepsilon.
\end{equation}
Write the continuity equation for this one-dimensional system in
the form

\begin{equation}\label{1dcontinuity}
    \left( H+\delta\right) v_1=\left( H+\varepsilon\right)
\widetilde{v_1}.
\end{equation}
Equations (\ref{conserve1}) and (\ref{1dcontinuity}) can be
combined into the single expression
\[
\left( H^{2}+2H\varepsilon+\varepsilon^{2}\right)
2g\varepsilon=v_1^{2}\left( 2H+\varepsilon+\delta\right) \left(
\varepsilon-\delta\right).
\]
Approximating this expression to first order in $\varepsilon$ and
$\delta,$ we obtain
\[
\varepsilon=\frac{\delta}{1-gH/v_1^{2}}=\frac{\delta}{1-\left(
c/v_1\right) ^{2}}.
\]
Again we find that the character of the flow is determined by
whether the Froude number $F$ exceeds, equals, or is exceeded by
the number 1, where in this case $\sqrt{Q}=\left| v_1\right| .$
The blow-up singularity that develops as $F$ tends to 1 \ is
avoided by hypothesis, as $\varepsilon$ is assumed to be small. In
the case of tranquil flow, a positive elevation $\left|
\varepsilon \right| $\ occurs when $\delta$ is exceeded by zero
and a negative elevation $-\left| \varepsilon\right| $\ occurs
when $\delta$ exceeds zero. The opposite relations hold for
shooting flow. Of course the effects of turbulence are ignored.

\section{A geometric variational problem}

Equation (\ref{flow}) can be derived by a variational principle
from an energy functional having the form

\begin{equation}\label{flowenergy}
    E_\rho\left(u;\Sigma,
\mathbb{R}^n\right)=\int_{\Sigma}\int_{0}^{Q}\rho( s) dsd\Sigma,
\end{equation}
where $\Sigma$ is a surface and $\rho(Q) =c^{2}.$ In generalizing
this problem we consider an energy functional of the form

\begin{equation}\label{energy1}
    E_\rho\left(u;M,N\right)=\int_{M}\int_{0}^{Q(du)}\rho\left(  s\right)  dsdM,
\end{equation}
where $M$ is a Riemannian manifold of dimension $n;$ $N$ is a
Riemannian manifold of dimension $m;$ $u:M \rightarrow N$ is a
bounded map;

\begin{equation}\label{energy2}
    Q\left(  du\right)  =\left\langle du,du\right\rangle _{\mid
T^{\ast}M\otimes u^{-1}TN};
\end{equation}
$\rho:\mathbb{R}^{+}\cup\left\{  0\right\}
\rightarrow\mathbb{R}^{+}$ is a $C^{1,\alpha}$ function of $Q\
$satisfying the differential inequality

\begin{equation}\label{subsonic}
0<\frac{\frac{d}{dQ}\left[  Q\rho^{2}(Q)\right]  }{\rho\left(
Q\right) }<\infty
\end{equation}
for $Q\in\lbrack0,Q_{crit});$ $Q_{crit}$ is the square of the
sonic flow speed. Inequality (\ref{subsonic}) is a condition for
tranquil channel flow.

We are interested in maps $u$ which are \emph{stationary} with
respect to $E_\rho\left(u;M,N\right)$ in the sense that

\begin{equation}\label{stationary}
    \delta E = \frac{d}{dt}E(u_t;M,N)|_{t=0}=0,
\end{equation}
where $u_t:M\times (-\epsilon < t < \epsilon)\rightarrow N ,$
$u_0=u,$ is a smooth, compactly supported one-parameter family of
variations (to be further specified below).

The variational equations of $E_\rho\left(u;M,N\right)$ are
satisfied by maps which are extremal within a competing homotopy
class of finite-energy maps from $M$ to $N.$ However, there are
solutions of (\ref{stationary}) which are not extremal with
respect to any class of maps.

Condition (\ref{subsonic}) is a condition for ellipticity of the
variational equation for $E,$

\begin{equation}\label{nhodge}
    trace\nabla_{cov}\left(\rho(Q)du\right)=0,
\end{equation}
where $\nabla_{cov}$ denotes the covariant derivative in the
bundle $T^{\ast}M\otimes u^{-1}TN.$ Note that this equation
introduces geometry into both the domain and range of eq.\
(\ref{flow}).

As an alternative to the hydrodynamic interpretation, the manifold
$N$ may be chosen to represent a geometric constraint placed on
the flow potential of a steady, irrotational, polytropic, perfect,
compressible fluid, which is adiabatic and isentropic and for
which the closed 1-form $du\in \Gamma\left(T^{\ast}M\right)$ is
dual to the flow velocity. \ In that case we choose
\begin{equation}\label{density}
\rho\left(  Q\right)  =\left(  1-\frac{\gamma-1}{2}Q\right)
^{1/\left( \gamma-1\right)  },
\end{equation}
where $\gamma>1$ is the adiabatic constant of the medium
\cite{Be}. Those choices transform (\ref{subsonic}) into a
condition for subsonic compressible flow of mass density $\rho.$
The \emph{sonic transition} as $Q$ tends to $Q_{crit}$ is a
gas-dynamic analogy for the change in the aspect of tap water from
clear to white at a sufficiently high velocity or for the
transition from tranquil to shooting channel flow in hydraulics.
We recover harmonic maps as the incompressible limit $\rho(Q)
\equiv 1.$

Note that two distinct terms are both referred to as
\emph{density} in the mathematical/fluid dynamics literature. The
physical, or mass density of the flow is $\rho(Q),$ but the
density of the variational integral $E_\rho$ is given by

\[
e(u)=\int_0^Q\rho(s)ds.
\]
In particular, the physical density given by (\ref{density}) is a
decreasing function of $Q,$ whereas the corresponding variational
density is an increasing function of $Q$ provided $Q <
2/(\gamma-1).$ The variational density $e(u)$ corresponding to the
physical density (\ref{density}) vanishes (or \emph{cavitates}) at
the flow speed $Q=0,$ but the physical density itself does not.
The physical density (\ref{density}) cavitates at the flow speed
$Q=2/(\gamma - 1),$ but the variational density does not.

Although in realistic physical contexts the ratio $\gamma$ of
specific heats must be taken to exceed 1, many of the analytic
properties of formula (\ref{density}) extend to the limiting case
in which $\gamma$ tends to unity. In this limiting case the
variational density is given by the function

\[
 \frac{1}{2}e(u)=1-\exp\left[-Q/2\right].
 \]
(We can easily see this by writing $y=\log \rho$ and applying
L'H\^opital's rule to (\ref{density}).) Regularity arguments for
weak subsonic solutions of (\ref{stationary}) presented for the
unconstrained case in, \emph{e.g.}, \cite{Sh} can be extended to
the limiting value of $\gamma$ by the Arzel\'a-Ascoli Theorem. One
thus obtains the existence of continuous subsonic solutions in
that limit by semicontinuity, using the convexity of the limiting
energy on the subsonic range (\emph{c.f.} Proposition 1 of
\cite{O1}). Some comments on the geometry of the limiting case are
given in Sec.\ 2.1 of \cite{O2}.

Reference \cite{O3} includes a review the isometric embedding of
the target manifold for critical points of (\ref{energy1}),
(\ref{energy2}) into a higher-dimensional Euclidean space
$\mathbb{R}^k$ and the use of nearest-point projection to obtain a
form of the geometric constraint which is convenient for
variational analysis in a Sobolev space. This is a familiar trick
in the theory of harmonic maps \cite{Sc}. We obtain variations of
the form $u_t=\pi_N\circ\left(u+t\psi\right),$ where $\psi$ is a
smooth map from $M$ into $\mathbb{R}^k$ and $\pi_N$ is the
\emph{nearest-point projection,} assigning to every $y$ in a
Euclidean neighborhood of $N$ the point on $N$ that minimizes the
distance to $y.$ The embedding of the flow geometry in Euclidean
space does not necessarily embed the corresponding physics in an
ambient Euclidean space. To illustrate the distinction, compare
water poured over a small sphere on the surface of Earth with
water flowing on the surface of a spherical planet. In the former
case we would take the gravitational acceleration vector to point
in the vertical direction of the Euclidean coordinate system in
which the sphere sits; in the latter case we would take the
gravitational acceleration to point in the radial direction of the
sphere itself. In the former case the gravitational potential
comes from the Euclidean space in which the sphere is embedded,
whereas in the latter case the intrinsic geometry of the surface
affects the gravitational potential.

A brief review of the literature relevant to eq.\ (\ref{nhodge})
$-$ in particular, its relation to the nonlinear Hodge theory
introduced in \cite{SS} $-$ is given in Sec.\ 1 of \cite{O3}. To
those remarks we add that the extension of geometric variational
problems for the Dirichlet energy to more general classes of
energies was already outlined in \cite{ES}, with a suggested
application of the harmonic map energy to the theory of elasticity
and that, whereas nonlinear Hodge theory generally considers the
elliptic case of the variational equations, eq.\ (\ref{nhodge})
will be allowed to change from elliptic to hyperbolic type as
$Q_{crit}$ is exceeded.

\section{Conditions for trivial flow}

We will say that a flow is \emph{trivial} if its flow potential
$u$ is a constant function. Equivalently, the velocity field of a
trivial flow associates the zero vector to every point of the flow
domain.

We expect that a relative minimum for a smooth function of a
single variable will occur at a point for which the second
derivative is non-negative. The analogue of the second derivative
for variational integrals is the \emph{second variation.} Let
$u_{s,t}:M \rightarrow N,$ $-\epsilon <s, T < \epsilon,$ be a
compactly supported two-parameter variation such that $u_{0,0}=u.$
Define

\[
V=\frac{\partial u_{s,t}}{\partial t}|_{s,t=0}
\]
and

\[
W=\frac{\partial u_{s,t}}{\partial s}|_{s,t=0}.
\]
The \emph{second variation} is the quantity

\[
I(V,W)=\frac{\partial^2}{\partial s\partial t}
E(u_{s,t})|_{s,t=0}.
\]
A map is said to be \emph{stable} if $I(V,V)$ is non-negative for
any compactly supported vector field $V$ along $u.$

This definition implies the triviality of a certain class of
flows:

\bigskip

\textbf{Theorem 1}. \emph{Let the domain of a shallow, steady,
irrotational hydrodynamic flow satisfying the hydrostatic law be
represented by a compact Riemannian manifold $M$ and let the flow
potential $u$ take $M$ into the $m$-sphere $\mathbb{S}^m$ for $m
\geq 2.$ Let the flow speed $Q$ be given by (\ref{energy2}) for
$0\leq Q<2.$ Then any stable flow potential takes every point of
$M$ to a single point on $\mathbb{S}^m$.}

\bigskip

\emph{Proof}. We impose a geometric constraint on the variational
problem (\ref{flow}), (\ref{flowenergy}), declaring that the image
of the flow potential $u$ must lie on a smooth, compact,
Riemannian manifold $N,$ where $N$ is a submanifold of
$\mathbb{R}^k$ for some sufficiently large number $k.$ This
transforms (\ref{flowenergy}) into (\ref{energy1}) and the
Euler-Lagrange equation (\ref{flow}) into the equation

\begin{equation}\label{flowalt}
    \delta_g\left[\left(1-\frac{Q}{2}\right)du\right]=\left(1-\frac{Q}{2}\right)A\left(du,du\right),
\end{equation}
where $\delta_g$ is the formal adjoint of the exterior derivative;
$g$ is the Riemannian metric on $M;$ $Q$ satisfies
(\ref{energy2}); $A$ is the second fundamental form of $N,$ where
$N$ is expressed as a submanifold of some higher-dimensional
Euclidean space as noted earlier. The system (\ref{flowalt}) is
identical to the variational equations for
$E_\rho\left(u;M,N\right)$ with $\rho$ given by (\ref{density}) in
the special case $C=n=\gamma=2,$ where $C$ is the constant of
Bernoulli's formula.

Computing the second variation of
$E_\rho\left(u;M,\mathbb{S}^m\right),$ we obtain

\begin{equation}\label{ineq}
    I(V,V)=\int_M |du|^2\left\lbrace |du|^2
\frac{d^2}{dQ^2}\int_0^Q\rho(s)\,ds+(2-m)\frac{d}{dQ}\int_0^Q\rho(s)\,ds\right\rbrace
dM.
\end{equation}
Because $\rho(s)=1-s/2,$ the right-hand side of (\ref{ineq}) is
negative under the hypotheses of the theorem unless $u$ is trivial
almost everywhere. But $u$ is continuous because $|du|$ is
bounded, so the conclusion holds everywhere on $M.$ This completes
the proof.

\bigskip

\textbf{Remarks}. \emph{i)} The conditions on the domain $M$ will
only correspond locally to the physical model of Sec.\ 1.2.

\emph{ii)} Computing (\ref{subsonic}) for $\rho$ given by
(\ref{density}) with $C=n=\gamma=2,$ we find that Theorem 1 holds
for flow speeds that extend well into the range of shooting flow.
Moreover, solving $Q=c^2=1-Q/2$ for $Q,$ we verify that the Froude
number attains the value 1 at $Q=2/3=2/(\gamma + 1)=Q_{crit}.$

\emph{iii)} A map $u:M \rightarrow N$ is said to be
\emph{$F$-harmonic} if it is a critical point of the $F$-energy
functional

\[
E_F(u;M,N) = \int_M F\left(\frac{|du|^2}{2}\right) dM,
\]
where $F:[0,\infty)\rightarrow [0,\infty)$ is a strictly
increasing, twice-differentiable function of its argument.
$F$-harmonic maps were introduced in  \cite{A1} as a unification
of $p$-harmonic and exponentially harmonic maps. A version of
inequality (\ref{ineq}) holds for any map which is $F$-harmonic;
\emph{c.f.} Theorem 7.1 of \cite{A1}, taking

\[
F\left(\frac{|du|^2}{2}\right)=e(u)=\int_0^{2\left(|du|^2/2\right)}\rho(s)\,ds.
\]
In fact, the extension of Theorem 1 to a larger class of target
manifolds is easily obtained by adapting the ideas of \cite{A2}.
However, Theorem 1 is false for the best known special cases of
$F$-harmonic maps: $p$-harmonic, exponentially harmonic, and
$\alpha$-harmonic maps; \emph{c.f.} the remark following Corollary
7.2 of \cite{A1}. Moreover, critical points of (\ref{energy1}),
(\ref{energy2}) will not be $F$-harmonic whenever the variational
density $e(u)$ is a decreasing function of $Q.$ That possibility
is allowed by condition (\ref{subsonic}) and by our initial
definition of $\rho.$

\emph{iv)} Theorem 1 recalls the famous statement that the wind
can never be blowing simultaneously in the same direction
everywhere on Earth. But of course the two assertions are
mathematically different and have completely different proofs.
They each result from the combination of a harsh global hypothesis
$-$ maximal symmetry in the range of a map $-$ with the global
imposition of what would be a reasonable local hypothesis $-$ in
one case, continuity and in the other, stability.

\emph{v)} Adding to the mass density $\rho(s)$ a generalized
``surface tension" of the form

\[
\tau(s)=\mu\left(1+s\right)^{-1/2},
\]
where $\mu$ is a positive constant, does not affect the result of
Theorem 1. But this extension is dependent on the sign of $\mu.$

\emph{vi)} Because the proof of Theorem 1 relies on applying a
compressible model to shallow hydrodynamic flow, it can be viewed
as a corollary of a theorem about compressible flow:

\bigskip

\textbf{Theorem 2}. \emph{Let the compact Riemannian manifold $M$
be the domain of a steady, polytropic, irrotational, perfect flow
and let the flow potential $u$ take $M$ into the $m$-sphere
$\mathbb{S}^m$ for $m \geq 2.$ Let the flow speed $Q$ be given by
(\ref{energy2}) for $0<Q<2/(\gamma-1),$ where $\gamma$ is the
adiabatic constant of the fluid. Then any stable flow takes every
point of $M$ into a single point of $\mathbb{S}^m.$}

\bigskip

\emph{Proof}. Follow the proof of Theorem 1 based on eq.\
(\ref{flow}) (\emph{c.f.} (2.14) of \cite{Be}), defining $\rho$ as
in (\ref{density}) for any $\gamma>1$ and any $n\geq 2.$

\bigskip

The correspondence between gas dynamics and shallow hydrodynamic
flow, illustrated in (\ref{flow}) and leading to the similarity of
Theorems 1 and 2, was apparently first reported in \cite{R} for
time-dependent compressible flow in 1 space dimension.
Elliptic-hyperbolic systems similar $-$ or dual $-$ to
(\ref{continuity2}) with a side condition of vanishing curl arise
in projective geometry and optics, as well as in hydrodynamics and
gas dynamics; \emph{c.f.} \cite{O4}.

This discussion provides a context for reinterpreting an earlier
result. The variational arguments in \cite{O} were given in terms
of a function $w\left(|du|^p\right).$ That function can be
interpreted as the variational density of the map $u,$ taking
$p=2$ and

\[
w(t)=\int_0^t \rho(s)ds.
\]
Thus the results of \cite{O} can be applied to hydrodynamic and
compressible flow, which gives another set of conditions for
triviality of the flow potential.

\end{document}